\documentclass[11pt]{article}
\usepackage{moriond,epsfig}
\usepackage{axodraw}
\usepackage{epsfig}

\def\cO#1{{\cal{O}}\left(#1\right)}
\def\eqref#1{eq.~(\ref{#1})}

\begin{document}
\vspace*{4cm}
\title{EFFECTIVE THEORY APPROACH TO UNSTABLE PARTICLES~\footnote{Talk given at XXXIX Rencontres de Moriond on Electroweak and Unified Interactions.}}% PRODUCTION}

\author{G. ZANDERIGHI}

\address{
Theoretical Physics Department, Fermi National Accelerator Laboratory, Batavia, IL 60510
}

\maketitle\abstracts{
We present a novel treatment of resonant massive particles appearing
as intermediate states in high energy collisions.  The approach uses
effective field theory methods to treat consistently the instability
of the intermediate resonant state.  As a result gauge invariance is
respected in every step and calculations can in principle be extended
to all orders in perturbation theory, the only practical limitation in
going to higher orders being the standard difficulties related to
multi-loop integrals.  We believe that the longstanding problem
related to the treatment of instability of particles is now solved.}

\section{The problem with unstable particles}
Physical phenomena studied in many ongoing and future collider
experiments involve the production of heavy unstable particles, such
as the top quark $t$, the vector bosons $Z^0, W^{\pm}$, the Higgs $H$
and, eventually, a range of particles appearing in extensions of the
Standard Model.

One powerful way to determine with high accuracy properties of these
particles (i.e.\ the mass $M$ and the width $\Gamma$) is to consider
their resonant production.  However, if treated as stable, unstable
particles give rise to non-integrable divergences in cross sections
due to the internal, divergent propagator $i/[q^2-M^2+i\epsilon]$
(with $q$ the four-momentum of the unstable particle).
A Dyson-summation of the self-energy $\Pi(q, M)$~\cite{Dyson}
\begin{equation}
\label{eq:dysonres}
\frac{i}{(q^2-M^2)+i\epsilon} \Rightarrow 
\frac{i}{(q^2-M^2)+i\epsilon}\sum_{n} 
\left(\frac{\Pi(q,M)}{q^2-M^2}\right)^n =  
\frac{i}{q^2-M^2- \Pi(q,M)}\>,
\end{equation}
ensures that the position of the divergence is away from the real
axis, since, due to unitarity, $\mbox{Im}\{\Pi(q,M)\}\ne 0$.
In the neighborhood of $q^2=M^2$, where $|\Pi(q,M)/{(q^2-M^2)}| >1$,
the series is strictly not convergent.  Eq.~(\ref{eq:dysonres})
assumes the validity of Perturbation Theory (PT) in the whole region
and uses analytical continuation in the problematic domain.

However, eq.~(\ref{eq:dysonres}) resums only a specific class of
higher order corrections, namely self-energy diagrams, while other
loop corrections, such as vertex corrections and box diagrams are
neglected completely. As a consequence, the resummation procedure
jeopardizes all properties valid order by order in PT, in particular
gauge invariance.
Many calculational schemes have been proposed in the last years to
overcome this longstanding problem (e.\ g.\ the fixed width scheme,
the running width scheme, the complex mass scheme~\cite{complexmass},
the fermion loop scheme~\cite{fls} and the pole
approximation~\cite{pole-scheme,ww-calc}).
%\cite{fls,complex-mass,pole-scheme,ww-calc}
%
The fundamental problem of most of these approaches is that in order
to restore gauge invariance physical quantities are modified ad-hoc,
introducing a certain degree of arbitrariness.
%, alternatively the pole
%approximation suffers from the limitation that it is not know how to
%go systematically beyond it. 
Furthermore it is not understood how to
improve the accuracy of predictions within these frameworks.
It turns out that in many cases the size of gauge violating terms is
numerically small and that various schemes do give quite similar
numerical results (though some approaches are known to give unreliable
predictions at higher center-of-mass energies~\cite{Denner:1999gp}).
However it is important to bear in mind that gauge dependent
theoretical predictions have ambiguities which can be in principle
arbitrarily large if enhanced by suitable kinematical factors or if
one chooses some pathological gauge.  It is therefore mandatory for
any precision measurement involving resonant unstable particles to
develop a framework, relying on first principles only, which allows
calculations to be performed to the required accuracy while preserving
gauge invariance.  This was the aim of the work presented here.

\section{The model}
To study the conceptual problem of instability it is convenient to
consider a simple model which allows one to avoid all unnecessary
technical difficulties, such as a large number of diagrams, integrals
containing multiple scales, or the presence of different interactions
whose coupling constants are numerically different, this would just
involve a subtler power counting.

The toy model that we consider involves therefore only an unstable
massive charged scalar field, $\phi$, two massless fermion fields, a
charged ``electron'' $\psi$ and a neutral ``neutrino'' $\chi$ and
finally an abelian gauge field, the photon $A^\mu$. The scalar decays
via Yukawa interaction into an electron-neutrino pair. The Lagrangian
describing this model is
\begin{eqnarray}
\label{eq:Lfull}
{\cal L} &=& (D_\mu\phi)^\dagger D^\mu\phi - \hat M^2 \phi^\dagger\phi +
 \bar\psi i \!\not\!\!D\psi + \bar\chi i\!\!\not\!\partial\chi
 - \, \frac{1}{4} F^{\mu\nu}F_{\mu\nu}-\frac{1}{2\xi} \,
(\partial_\mu A^\mu)^2
 \nonumber\\
 && + \, y\phi\bar\psi\chi + y^* \phi^\dagger \bar\chi\psi
-\frac{\lambda}{4}(\phi^\dagger \phi)^2+ {\cal L}_{\rm ct}\, ,
\end{eqnarray}
where $\hat{M}$ and ${\cal L}_{\rm ct}$ denote the renormalized mass
and the counterterm Lagrangian respectively and $D_\mu=\partial_\mu-i
g A_\mu$ is the covariant derivative. We define $\alpha_g\equiv
g^2/(4\pi)$, $\alpha_y\equiv (y y^*)/(4 \pi)$ (at the renormalization
scale $\mu$) and assume $\alpha_g \sim \alpha_y \sim \alpha$, and
$\alpha_\lambda\equiv\lambda/(4\pi) \sim \alpha^2/(4\pi)$.

In the following we will discuss the elements needed to compute the
inclusive cross section for
\begin{equation}
\bar\nu(p_1) + e^-(p_2)\to X\,,
\label{process}
\end{equation}
as a function of $q^2\equiv (p_1+p_2)^2$ in the resonant region, where
$\delta\equiv(q^2-M^2)/q^2\sim \alpha\sim \Gamma/\hat M$.
To illustrate the numerics we will use for the
$\overline{\mbox{MS}}$-mass $M=100$GeV and for the couplings
$\alpha_y=\alpha_g=0.1$ and $\alpha_\lambda=(0.1)^2/(4\pi)$, evaluated
at the renormalization scale $\mu = M$.

\section{The effective theory}
It is well known that in scattering processes involving an unstable
particle, radiative corrections can be split into {\it factorizable}
and {\em non-factorizable} corrections.~\cite{Khoze:1992rq} The latter
are quantum fluctuations which have a propagation time comparable to
the one of the unstable particle $\tau \sim 1/\Gamma$, while the
propagation time $t \sim {1}/{M} \ll \tau$ of the factorizable
corrections does not allow any quantum interference with the unstable
field.
Factorizable corrections are therefore hard loop effects, while
non-factorizable corrections are described by fluctuations of soft
($p_s\sim \delta M$) and collinear ($p_c \sim M, p_c^2 \sim M \delta$)
massless fields or resonant unstable massive particles ($q^2-M^2\sim
\delta M$).
The separation between hard and soft/collinear is manifestly gauge
invariant and can be performed to all orders in PT using an effective
field theory approach,~\cite{SCVK} rather than a diagrammatic one.
This classification of radiative corrections can thus be used to
construct an effective theory by integrating out the hard degrees of
freedom which end up in the Wilson coefficients of the effective
operators, while soft, resonant or collinear fluctuations remain
dynamical modes of the effective theory.

Since all fields have a well-defined scaling in $\delta$, the
contributions from different regions can be selected with the strategy
of regions,~\cite{Regions} which gives an unique prescription of how
to expand any given loop integral in powers of $\delta$.
Dimensional regularization ensures than that regions with no
singularities do not contribute, avoiding the problem of
double-counting.
Of course, by splitting an integral into different regions one might
generate spurious singularities. However, the infrared singularities
of the hard fluctuations cancels against the ultraviolet singularities
of the dynamical modes. We choose to subtract them minimally hereby
defining a renormalization scheme in the effective theory.

The construction of the effective theory proceeds along the standard
way. Let us therefore only review the main steps. All technical
details are given elsewhere.~\cite{BCSZ} Similarly to Heavy Quark
Effective Theory (HQET)~\cite{HQET} we first redefine the massive
field, we extract the big momentum component, so that derivatives
acting on the new field $\phi_v$ produce only soft fluctuations
\begin{equation}
\phi_v \equiv e^{i \hat M vx} {\cal P}_+\phi\,,\qquad 
\partial\,\phi_v = \cO{\delta}\, \phi_v\>.  
\end{equation}
Here ${\cal P}_+$ projects out the positive frequency states so that
$\phi_v$ is a pure annihilation operator.  
The actual matching procedure is based on requiring that on-shell
Greens functions in full theory and in the effective theory coincide
up to the required accuracy.

In the resonant region, the effective Lagrangian resulting
from~\eqref{eq:Lfull} after integrating out the hard modes, is given
by a sum of three physically distinct contributions.  The propagation
of the unstable particle gives rise to terms in the effective
Lagrangian which closely correspond to the well-known HQET Lagrangian
\begin{eqnarray}
\label{eq:HSET}
{\cal L}_{\rm HSET} &=&  2 \hat{M}  \phi_v^\dagger\, 
        \left( i v \cdot D_s - \frac{\Delta^{(1)}}{2} \right) \phi_v 
+ 2 \hat{M}  \phi_v^\dagger\,
        \left( \frac{(i D_{s,\top})^2}{2\hat{M}} +
               \frac{[\Delta^{(1)}]^2}{8\hat{M}} -
               \frac{\Delta^{(2)}}{2} \right) \phi_v 
\nonumber  \\
&-&\frac{1}{4}\,F_{s\mu\nu} F_s^{\mu\nu}
+\bar\psi_s i\!\not\!\!D_s \psi_s
+\bar\chi_s i\!\not\!\partial \chi_s\,,
\label{eq:Leff_phi}
\end{eqnarray}
where the subscript $s$ indicates that only the contribution due to
soft fields has been included, the subscript $\top$ indicates the
transverse direction with respect to the velocity vector of the
unstable field. $\Delta^{(i)} = \cO{\alpha^i}$ are matching
coefficients which are defined by
\begin{equation}
\sum_i \Delta^{(i)} \equiv \frac{\bar s -\hat M^2}{\hat M^2}\,,
\end{equation}
with $\bar s$ the position of the pole of the propagator, which 
can be computed from the analytic hard part of the self-energy.  
${\cal L}_{\rm HSET}$ describes hence the interaction of the unstable
particle with soft gauge fields and soft fermions.

The terms in \eqref{eq:Lfull} describing the interaction of energetic
fermions in the initial and final state and gauge fields give rise to
the well-known soft-collinear effective theory,\cite{SCET}
\begin{equation}
\label{eq:Leff_SCET}
{\cal L}_{\rm SCET} =
\bar{\psi}_c \left(i n_- D +  i \!\not\!\!D_{\perp c}
\frac{1}{i n_+ D_{c}+i\epsilon}\, i\!\not\!\!D_{\perp c} \right)
\frac{\not\!n_+}{2} \, \psi_c\, . 
\end{equation}
Here the subscript $c$ indicates contributions from collinear fields
along the $n_-$ direction ($n_-^2=n_+^2=0, n_-n_+=2$). Obviously, the
SCET Lagrangian for neutrinos can be obtained from
\eqref{eq:Leff_SCET} by replacing covariant derivatives with partial
ones.

Finally, the Lagrangian contains Yukawa vertexes which allow the
production and decay of the unstable field
\begin{equation}
{\cal L}_{\rm int} = C\, y\, \phi_v \bar{\psi}_{n_-}\chi_{n_+} 
+ C\, y^* \phi_v^\dagger \bar{\chi}_{n_+}\psi_{n_-} 
+ F\, y y^* \left(\bar{\psi}_{n_-} \chi_{n_+}\right)  \!
  \left(\bar{\chi}_{n_+} \psi_{n_-}\right)\,,
\label{eq:Lint}
\end{equation}
where $C=1+{\cal O}(\alpha)$ and $F$ are matching coefficients.
${\cal L}_{\rm int}$ involves new external-collinear fields
$\psi_{n_-/n_+}$ and $\chi_{n_-/n_+}$, whose momenta are given by $\hat
M/2n_{-/+} + k$, with $k=\cO{\delta}$. For instance $\psi_{n_-}$ is
defined as
\begin{equation}
\psi_{n_-} \equiv e^{i \hat M/2 (n_- x)}{\cal P}_+ \psi_c (x)\>. 
\end{equation}
The introduction of these fields is useful to distinguish situations
were two generic collinear fields produce to a state with invariant
mass of order $M$, from situations involving two external collinear
fields which give rise to states with invariant mass equal to $M$ up
to $\delta M$ corrections.

Once the effective Lagrangian has been established, one can compute
physical quantities using the gauge invariant formalism following from
it. In particular we will present here results for the inclusive
line-shape in our toy model. Since all fields in the effective theory
have well-defined scaling properties, a power-counting allows to
identify the operators that are needed to a given order and at which
order in the coupling they need to be matched.

The effective approach is designed to describe the problem in the
resonant region, where $\delta\sim\alpha$, therefore at leading order
(LO) we resum all terms $\cO{\left({\alpha}/{\delta}\right)^n}$. One
order beyond, at NLO, we include all terms
$\cO{\left({\alpha}/{\delta}\right)^n \delta}$.  NLO terms are
expected to correct the LO result by $\sim
\delta\sim 10\%$. In a similar way NNLO corrections resum
$\cO{\left({\alpha}/{\delta}\right)^n\delta^2}$ and are expected to be
of the order $\sim \delta^2\sim 1\%$.

At LO the matching amounts simply to resum in the propagator
{\em on-shell} self-energies. At NLO, one needs to match vertex
corrections at subleading order and to include sub-leading corrections
to the propagator. Furthermore dynamical corrections must be also
included.  

One of the main results of this work is the possibility to perform a
NNLO calculation of line-shapes.  Even if it is not sensible to
struggle with a NNLO calculation in a toy model, it is interesting to
see what elements are needed to achieve NNLO accuracy.  We use the
intuitive notation where $\alpha_h$ denotes an $\cO{\alpha}$
correction due to a hard loop (and similarly for soft/collinear
loops).  We can thus classify NNLO corrections according to
\begin{itemize}
\item tree amplitude in the effective theory with 
insertions of LO operators in $\delta$, matched to NNLO
($\alpha_h^2$); 
\item tree amplitude in the effective theory with 
insertions of NLO operators in $\delta$, matched to NLO
($\alpha_h\delta$);
\item tree amplitude in the effective theory with 
insertions of NNLO operators in $\delta$, matched to LO ($\delta^2$); 
\item one-loop amplitude in the effective theory with 
insertions of LO operators in $\delta$, matched to NLO
($\alpha_s\alpha_h,\alpha_c\alpha_h$);
\item one-loop amplitude in the effective theory with 
insertions of NLO operators in $\delta$, matched to LO 
($\alpha_s\delta,\alpha_c\delta$); 
\item two-loop amplitude in the effective theory with 
insertions of LO operators in $\delta$, matched to LO 
($\alpha_s^2,\alpha_s\alpha_c,\alpha_c^2$). 
\end{itemize}
This simple classification shows that NNLO calculations of line-shapes
are now technically feasible and that they can be carried out in a
systematical and transparent way.
We note also that due to different kinematics the various
contributions are separately gauge invariant.

\begin{center}
\begin{figure}[t]%[hpt]
\begin{minipage}{0.45\textwidth}
    \hspace{.8cm}\epsfig{file=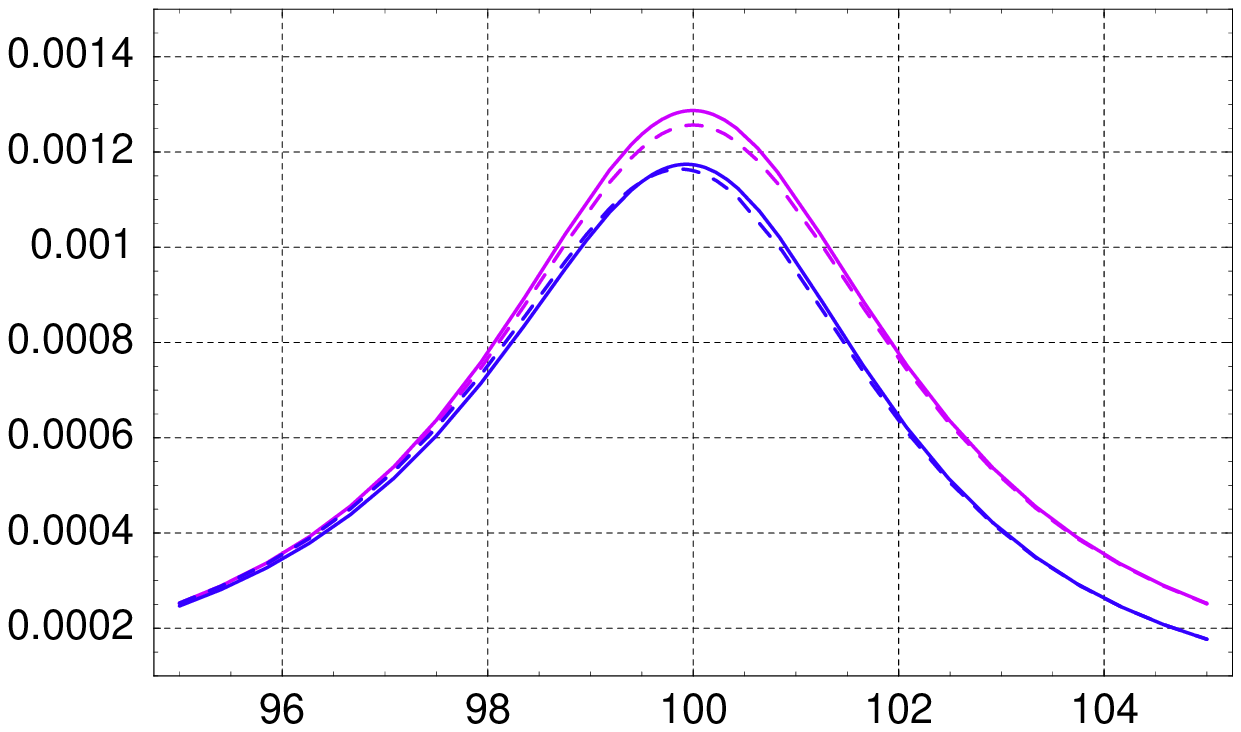, width=1.\textwidth, angle=0}
\end{minipage}
\begin{minipage}{0.45\textwidth}
\hspace{1.3cm}    \epsfig{file=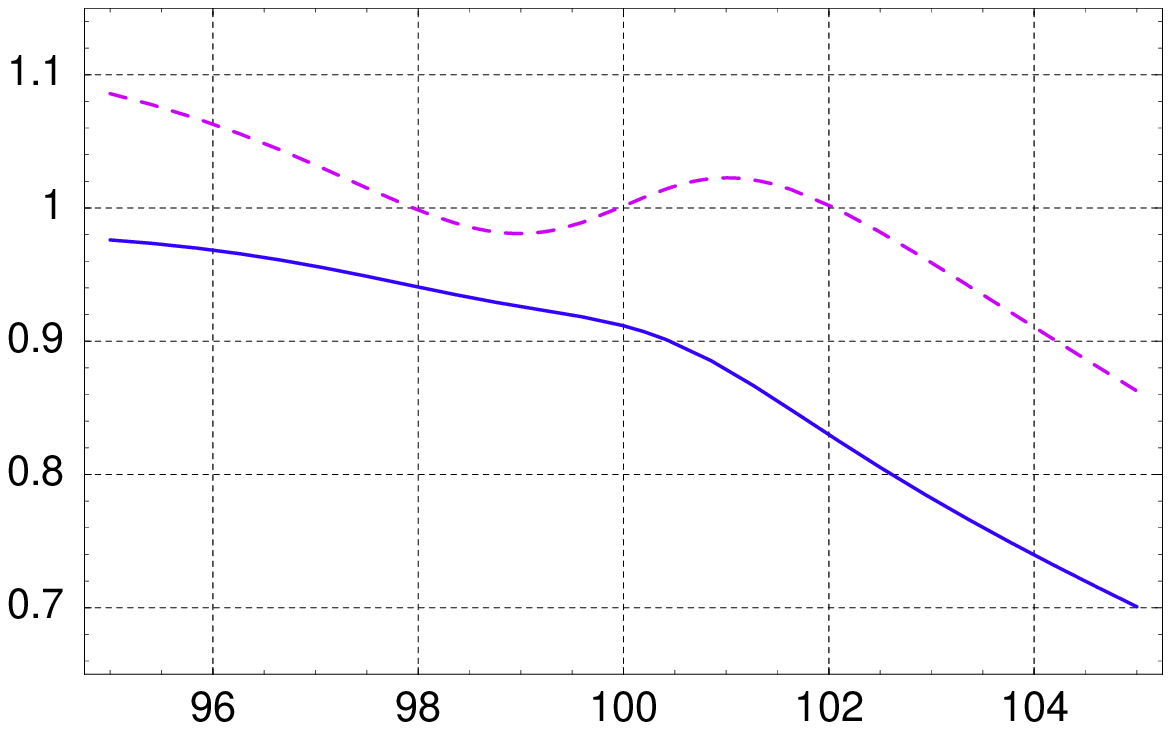, width=1.\textwidth, angle=0}
\end{minipage}
  \caption{
Left: line shape (in $\mbox{GeV}^{-2}$) in the 
  $\overline{\rm MS}$ scheme (solid) and pole scheme
  (dashed) at LO (upper magenta/light grey curves) and NLO
  (lower blue/dark grey curves) as a function of the center-of-mass 
  energy in GeV. 
  Right: The ratio of the NLO to the LO line shape in 
  the $\overline{\rm MS}$ scheme (solid blue/dark grey curve) 
  and the ratio of the NLO  $\overline{\rm MS}$ line shape to a 
  Breit-Wigner fit of the NLO result (dashed magenta/light grey curve).
}
\label{fig:sigmaNLO}
\end{figure}
\end{center}

We computed the total cross section from the forward scattering
amplitude.
Fig. \ref{fig:sigmaNLO} (left side) shows the LO and NLO line-shape
in the $\overline{\rm MS}$ scheme and pole scheme. The NLO correction
amounts to 10\% at the peak and to up to 30\% in the resonant region.
The right panel shows the ratio of the LO over the NLO line-shape.
In the absence of data we performed a fit of the NLO line-shape to a
Breit-Wigner and we plot the ratio of the NLO result to the
Breit-Wigner fit.
As can be nicely seen from Fig. \ref{fig:sigmaNLO} (right side) the
deviation between NLO and the Breit-Wigner amounts to up to 15\%.
Furthermore, the value of the mass fitted from the Breit-Wigner
differs from the input mass (the {\it true} value) by $160\,
\mbox{MeV}$.
This shows explicitly that for precision studies a proper theoretical
prediction should be used, rather than a Breit-Wigner fit, which
produce sizable shifts in the mass prediction.

%=====================================================================
\section{Concluding remarks}
The perturbative treatment of unstable particles requires a partial
summation of the perturbative series, however we believe that the
guiding principle was not understood. The breakdown of PT is related
to the appearance of a second small parameter $\delta$, besides the
coupling $\alpha$. We take the attitude that this is {\it the}
characteristic feature of the problem, so that in a theory that
formulates this double-expansion correctly, other issues like
resummation, gauge invariance and unitarity should follow
automatically. Since it is a two-scale problem ($\Gamma \ll M$) an
effective theory is the natural framework to formulate such an
expansion.
The advantages of using effective theory methods are 
\begin{itemize}
\item calculations are split into well-defined pieces (matching, matrix
  elements, loop calculations in the effective theory \dots ), so that
  the calculation is efficient and transparent;
\item a power counting scheme in the small parameters
  ($\alpha$, $\delta$) allows one to identify terms required to
  achieve a certain accuracy prior to performing the actual calculation; 
\item the effective theory provides a set of simpler Feynman rules
  which allows one to compute the minimal set of terms required at the
  given accuracy. Since one never computes ``too much'', calculations
  are as simple as possible;
\item calculations can be extended in principle to any order in
  $\alpha, \delta$, at the price of performing complicated, but
  standard loop integrals;
\item since the expansion has been organized in
  such a way so as to account for kinematical enhancements the PT
  series in the effective theory converges rapidly;
\item gauge invariance is automatic. 
\end{itemize}
Despite the simplicity of the model considered (abelian theory, scalar
particle), all necessary ingredients are provided for the formalism to
be applied to any general case. Natural extensions concern
non-inclusive kinematics, which requires a formalism to expand the
real phase-space and generally implies that more collinear directions
are relevant, and to pair-production of unstable particles, in which
case the effective Lagrangian will contain two terms similar to
\eqref{eq:HSET} describing the propagation of the two particles with
different velocity vectors.

\section*{Acknowledgments}
I thank Martin Beneke, Sasha Chapovsky and Adrian Signer.  During this
enjoyable and fruitful collaboration I learned a lot from each of
them.  I'm also grateful to Andrea Banfi and Uli Haisch for useful
comments and careful proofreading of the manuscript. This work is
supported by the U.S. Department of Energy under contract
No. DE-AC02-76CH03000.

\end{document}